\documentstyle[12pt,aasms4]{article}

\begin{document}

\title{LABORATORY  SYNTHESIS  OF  MOLECULAR  HYDROGEN  ON   SURFACES  OF
ASTROPHYSICAL   INTEREST }
\author{V.Pirronello}
\affil{Istituto   di    Fisica, 
Universita' di Catania, Catania,  Sicily,  Italy 
and the Department of Physics, Syracuse University
Syracuse, NY 13244-1130}
\author
{Chi Liu, Liyong Shen, and Gianfranco Vidali $^{*}$}
\affil{The Solid State Science and Technology Program 
and the Department of Physics, Syracuse University,
Syracuse, NY 13244-1130}

\begin{abstract}
We report on
the first results of experiments to measure the recombination
rate of hydrogen on surfaces of astrophysical interest. Our
measurements give lower values for the recombination efficiency
(sticking probability $S$ x probability of recombination upon H-H
encounter $\gamma$)
than model-based estimates.  We
propose that our results can be reconciled with {\it average}
estimates of the recombination rate ($1/2 n_H n_g v_H A S
\gamma$)
from astronomical observations, if the actual surface of an average grain is
rougher, and its area bigger, than the one considered in models.
\end{abstract}

\section{Introduction}

$H_2$ formation in the ISM (interstellar medium) is a fundamental
process in astrophysics.
The radiative association of two hydrogen 
atoms is a process too rare to be efficient because it involves
forbidden roto-vibrational transitions, and gas-phase three-body
reactions are rare in the diffuse ISM 
to explain $H_2$ abundance.  It has long been
recognized that hydrogen recombination occurs on surfaces of dust
grains, where the grains  act as the third body in the $H+H$ reaction. 
(\cite{duley}).

The critical physical
steps in which interstellar grains act as catalysts for the $H_2$ synthesis
are: sticking and accommodation of gas-phase atoms on the grain
surface; mobility of H adatoms to meet each other within
their residence time on the cold grain; and ejection of the newly
formed molecule into the gas phase.

The efficiency of these processes depends on the
structure and the chemical activity of the grain surface, whether it is
silicon-bearing  (silicates can be not very stable in UV radiation fields)
\cite{hong,greenberg},
carbonaceous (carbon can possibly be distributed on silicon-bearing grain
surfaces) or icy
(\cite{mathis}).
In the case of more chemically active surfaces, like carbonaceous
ones, the possible presence of unsaturated bonds will increase the
percentage of chemisorption events. The morphology of the surface
should influence the sticking and tunneling processes; it is
reasonable to assume that sticking will be higher and diffusion length lower
on amorphous than on crystalline surfaces.  Amorphous is the most
probable structure for grains in space, as shown by observations
(\cite{leger}).

Observed $H_2$ abundance in diffuse clouds 
can be explained if the formation rate $R$ on grains balances its destruction
rate; the canonical value of $R$ is
$\sim 3 x 10^{-17} cm^{3} sec^{-1} $
(\cite{jura}); 
on grain surfaces, the recombination rate can be described as  $\sim 1/2 n_H n_g v_H A S
\gamma$ where, $n_H$ and $n_g$ are the number densities of hydrogen
atoms and dust grains, respectively, and  $v_H$ is their relative velocity. $S$
(sticking coefficient) is the probability that an atom hitting a dust
grain remains on the surface, and $\gamma$ is the probability that an
atom, once on the surface, makes an encounter with another H atom and recombines 
with it. For a
typical diffuse cloud environment, 
(\cite{jura})
it is assumed that, on average, $S \gamma \sim 0.3$.  

From a theoretical standpoint, the complexity of the problem lies in
the fact that, due to the quantum nature of the H atom, a quantum
mechanical calculation has to be done on a realistic, heterogeneous
surface resembling dust grains in ISM.
Hollenbach and Salpeter (HS)
(\cite{hollenbach}) 
calculated $S$ using  simple semiclassical method for the atom-surface
interaction, and found $S$ to be between
$10^{-1}$ and 1 in most cases. Leitch-Devlin and
Williams
(\cite{leitch})
used a quantum mechanical approach and assumed a perfect single
crystal with energy loss due to single phonon excitations. Their
sticking coefficient increases with gas temperature, reaches a maximum
at $ k_BT$ comparable with phonon energies and then decreases
again.  Buch and Zhang
(\cite{buch})
numerically evaluated the sticking of hydrogen atoms on a cluster
(amorphous in structure) of water molecules. They
obtained $ S= (k_BT/E_0 +1)^{-2}$, where $E_0$ is a parameter($\sim 100 k_BT$
for $H$ and $\sim 200 k_BT$ for $D$)
 
In HS's calculation 
(\cite{hollenbach}),
the tunneling efficiency of adsorbed hydrogen was high
enough to ensure a very fast scanning of virtually all possible
adsorption sites on a grain surface in a fraction of the residence
time of the adsorbed species. In their model, the grain
surface has two adsorption sites for H, one weak ($\sim 400 K$) and
the other strong ($\sim 20,000 K$). If only one weak
adsorption site is present, H atoms wouldn't spend enough time on the
surface to meet other H atoms and they would evaporate in a time $\tau
\sim \tau_0 exp(E/k_BT)$, where $\tau_0 $ is a typical time of a
vibration of a H atom in an adsorption site.

Smoluchowski 
(\cite{smol})
did a quantum mechanical calculation and found that $H$ would get trapped in the 
deepest sites of an amorphous
surface after a few 
hops. His efficiency of $H_2$ production is several orders of
magnitude smaller than HS's.  Pirronello and Averna
(\cite{pirro})
investigated the possibility  that $H_2$ could be
produced  in dense clouds by cosmic rays bombardment of grain mantles.
Within the  Smoluchowsky's model, they found that the 
cosmic rays mechanism would be dominant over $H_2$ recombination on
grains at 10 K.

Although sticking has been extensively studied experimentally  
(\cite{rend}), 
it has been done in conditions and on surfaces of almost no
astrophysical interest, except for the following studies.  Brackmann and Fite
(\cite{brackmann})
measured a sticking probability of 0.2 on a $H_2$-free surface and 0.5
on top of a $H_2$ layer in the temperature range of 2.5 to 8 K. Due to
poor vacuum conditions of their apparatus, it is assumed that their
surfaces were covered with thick cryodeposits of background gas.
There have been even fewer determinations in the laboratory of the hydrogen 
recombination coefficient on surfaces and up to now in conditions of no astrophysical
relevancy.  Schutte et al.
(\cite{schutte})
measured the H recombination on a surface of a bolometer at 3 K, a
temperature  significantly smaller than the one  of
interstellar grains (10-15 K). They found $S\gamma
\sim 0.05 - 0.1$ on a hypothetically $H_2$-free surface.
Other experiments
(\cite{schermann,king})
measured recombination rates in situations where either the
surface temperature or the kinetic energy of the H atoms were high. 
On the other hand, $H$ beam scattering experiments have shown that the
interaction of {\it low energy} $H$ beams with single crystal graphite
surfaces is rather weak; the ground state is $\sim32$ $meV$ deep.
(\cite{mattera})
Experiments of scattering of $H$ from graphite single crystals above
$\sim 16$ K gave a sticking coefficient of less than 0.1 and
negligibly small above 21 K.
(\cite{lin})

The experimental studies mentioned above, although gave
interesting information about physical/chemical process
at surfaces, didn't really address the
question of measuring the hydrogen recombination rate on surfaces in
situations which could be related to astronomical observations. Our
study attempts to address this need.

In this report, we present the results of a set of experiments to
measure the recombination coefficient on the surface of an olivine
slab (a silicate rich in Fe, Mg, Si and oxygen), an 
astrophysically relevant material, even if polycrystalline,  as 
natural stones on Earth are, and not amorphous, as required by 
observations
(\cite{draine}).
Low kinetic energy of H
beams, low fluxes, and low surface temperature  have been used.

\section{Experimental Set-Up}

The experiments were performed in an ultra-high vacuum apparatus
consisting of a scattering chamber and of two triply differentially
pumped beam lines (See Fig. 1). Each atomic beam is produced by RF
dissociation of 
molecular hydrogen (or deuterium) in a water-cooled Pyrex tube placed
in a RF cavity. An ENI power supply provides 100-150 watt of power to the
cavities via a power splitter and impedance matching networks. 
By using $H$ and $D$ beams, the recombination
product ($HD$) can be formed only on the surface, and not in the beam
source due to imperfect dissociation, as in the case when using only one
beam. The
dissociation rate is measured by the quadrupole mass spectrometer in
the main chamber. The maximum dissociation rate is over 90 $\%$, but 
in most cases was
lower, around 70-85 $\%$, and it remained constant
during experimental runs. Each beam is made by the expansion of a low
pressure gas (0.1-0.2 torr) into vacuum through a short aluminum
capillary of 1mm in diameter connected to a  $LN_2$
reservoir. The beams
enter the scattering chamber 
through 3 mm collimators. The estimated solid angle is $ 6 x 10^{-6}
sr$ and the flux for H is $\sim 10^{12}$ atoms/cm$^2$/sec. To obtain a
lower flux, a mechanical chopper with a duty cycle of $1:20$ was used.
The $H$ beam hits the 
surface of the sample perpendicularly, while the $D$ beam hits it at 38
deg. from the normal. The sample is mounted on a copper sample holder that
is in good
thermal contact with a HeliTrans cold finger and its temperature
is measured via two calibrated chromel/iron-gold thermocouples
pressed against the top and bottom surfaces of the sample. The lowest
temperature reached with the cold finger is
$\sim 5-6 K$ (top thermocouple). The sample can be heated by radiation  and
electron-beam bombardment with a heater placed in a small cavity of
the sample holder just behind the sample. The incoming and
reflected beams are detected by a differentially-pumped quadrupole
mass spectrometer. 

\section{Measuring Procedures and Results}

Most of the results reported here are for a sample of olivine donated
by Dr. P. Plescia of the CNR Institute for the 
Treatments of Minerals (Rome). It consists of a mixture of $Fe_2SiO_4$
and $Mg_2SiO_4$. Prior insertion into the apparatus, the sample was
cleaned with mild solvents (acetone, methanol, freon) in an ultrasonic
bath. The sample was then placed on the sample holder in the UHV
chamber and the apparatus was baked to 150 $^{\circ}$C for a couple of
days. The 
base pressure was in the mid $10^{-10}$ torr range.  In a typical
experiment, the sample is first heated to 200 $^{\circ}$C for cleaning. After
cooling, it is exposed for a given amount of time to the $H$ and $D$
beams.   During the adsorption time, the detector is placed in
front of the sample and the mass is tuned to 3 (i.e., mass of
$HD$. There is no other background gas contributing to the 
spectrometer signal at mass ``3''). At the end of the exposure time, the sample
temperature is quickly ($\sim 1$ K/sec) raised to over 30 K by
shutting the He flow to the tip of the cryostat; the amount of $HD$
desorbed as a 
function of time is recorded into a multichannel scaler. A set
of representative traces is shown in Fig. 2. An analysis of desorption
kinetics is given elsewhere. 
(\cite{liu})

The background pressure of $HD$ due to finite
pumping capacity and/or recombination of $H$ and $D$ on the walls of
the chamber is measured prior to and after each adsorption/desorption
experiment, in front and back of the sample, and is subtracted from the
recorded signal.  We checked 
that after the blow-off of the $HD$ layer there is no re-adsorption of $HD$
from the gas phase.  The sample holder is hugged by a copper radiation
shield; no adsorption/desorption of $HD$ has been detected on/from the
shield whose temperature is considerably higher
than the sample. 
Exposure times have been changed over several decades, see Fig. 3,
where the desorption yield (which is the integral of curve in Fig.2)
is plotted vs. exposure to $H$ and $D$. The data are fitted well
by a Langmuir adsorption kinetics. The saturation of the
signal is interpreted as due to the completion of the first layer.
The time it takes to form a $HD$ layer is consistent with the calculated
flux and a low ($\sim 0.1$) sticking coefficient
(\cite{brackmann,schutte}). 

Fig. 4 shows the amount of $HD$ desorbed following adsorption of $H$
and $D$ at various sample temperatures. The amount of $HD$ produced
and released 
{\it during} the adsorption process (as measured by the mass
spectrometer) is typically much smaller than the amount
released in the thermal desorption run. This means that at low sample
temperatures (5-7 K), $H$ sticks (although with a sticking probability
well below 1)
and readily recombines; because of the low temperature, just 
$HD$ that has just formed remains on the surface. At higher sample
temperatures (10-15 K),  $H$ doesn't stay on the
surface long enough to recombine, and the $HD$ yield is much
smaller. Likely, only those $H$ atoms which became trapped
in strong binding 
sites are retained at these temperatures; $HD$ is then produced at
reduced rates because of 
slow diffusion of $H$ atoms out of these deeper energy sites.
We couldn't measure the $H$ signal during
adsorption/desorption in a quantitative way because of the presence of
background of $H$ in a stainless steel system (even at 5 $10^{-10}$
torr total base 
pressure). An analysis of desorption kinetics after adsorption of $H$ at
high sample
temperature could yield some valuable clues; at present, however, the
desorption signals are too low for a meaningful quantitative analysis.

\section{Discussion}

The recombination efficiency of $HD$, here defined as the fraction of $H$
atoms which stick and recombine to from $HD$ - or $r\sim S\gamma$, is
calculated as 
follows (Note that this is different from the recombination rate
$R$ defined earlier):  $r = (I_a + I_d) / I_{in}$, where $I_a$ is the $HD$ signal
during adsorption, $I_d$ is the total amount of $ HD$ desorbed after
the adsorption process, and $I_{in}$ the the amount of H sent into the
chamber. The signals are corrected for velocity-dependent detection
efficiency and for the measured solid angle of the detector.$ r$ is also
corrected for the probability of forming $HD$
into other reaction channels, and for the different
measured intensities of the $H$ and $D$ beams. As mentioned before,
only a small fraction of the amount of $HD$ which desorbs in the
thermal programmed 
desorption leaves the sample during the {\it adsorption process}. 
Consequently, the
recombination efficiency drops with increasing
surface temperature similarly to the fall off of the desorption yield
as a function of temperature, see Fig. 4.  The recombination efficiency was
measured for different exposures to $H$ and $D$, as seen in Fig. 5.

Other candidate materials for grains in the ISM are:
carbonaceous solids, both amorphous and crystalline, and icy mantles
(\cite{mathis}). 
We plan to investigate the formation of molecular
hydrogen also on these other materials and we have already obtained
some
preliminar results (to be published elsewhere) for $D_2$
recombination on HOPG (Highly Oriented
Pyrolitic Graphite).

Our estimate of $r$ at 5 - 6 K is not far from Schutte et al.' s value
obtained on 
a uncharacterized surface of a semiconductor bolometer but at
considerably lower temperature (3K). At the astrophysically relevant
grain temperature range of 10-15 K, our  result, $r \sim 0.03 - 0.05$ is
noticeably lower than the value of HS's model 
(\cite{hollenbach}), 
$\sim 0.3$ at 10-15 K.

We might try to speculate on the reason of this difference.
Calculations are for very idealized surfaces. There are a few
assumptions that can influence the outcome of the calculation,
such as the semiclassical description, the energetics of the binding
sites, and the tunneling time between sites. The sample used by us
presents a far more heterogeneous environment, both energetically and
morphologically, than it has been considered in
theoretical models. In a real surface, tunneling might proceed much
more slowly, and the sticking coefficient for H on surfaces of
olivine and graphite might be much different than previously
calculated on model surfaces.
Considering that many models 
(\cite{duley})
assume $S\sim 0.3$, and that a
determination of $S$ for $H$ on graphite at $\sim 16 K$ gives $S\sim
0.06 - 0.1$
(\cite{lin})
, it is conceivable that $S$ is only partially responsible
for the fact that our $r=S\gamma$ is a factor 10 lower than
Hollenbach's, and that $\gamma$ might be appreciably less than $1$,
contrary to what is most often assumed in models. 

What are the consequences of this determination of $S \gamma$ on the
recombination 
rate $R$ giving the constraints posed by astronomical observations?
In order to reconcile our value for $r=S \gamma \sim 0.03 - 0.05$ with
the accepted value 
(\cite{jura}) 
for
$R$ ($\sim 1$ to $ 3 x 10^{-17} cm^{3} sec^{-1} $), we propose that
the surface area of grains (which, in a simple model, enters the
expression for $R\sim 1/2 n_H n_g v_H A S
\gamma$ as a multiplicative factor) might be larger (by a factor 5 to
10 or even more for amorphous surfaces). This could be indeed the case if 
grains have a larger surface area (i.e, they are ``fluffier')
(\cite{whiffen}),
than previously considered.
This is not an
unlikely scenario, considering 
the type of processing that grains are subjected to in the ISM, and more
theoretical and experimental effort should be devoted to verify it.

\vskip 0.3 true in \noindent{\large Acknowledgments}

\acknowledgments
Support from NASA-Astrophysics Division is gratefully
acknowledged. We thank Dr. PLescia of the CNR - Rome for providing the
olivine sample and  ENI of Rochester, N.Y. for donating the RF splitter.
\vskip 0.3 true in \noindent
$^{*}$ To whom correspondence should be addressed; e-mail: gvidali@syr.edu
\vspace{0.5in}

\newpage
 

\figcaption{Schematic top-view of the apparatus}
 
\figcaption{Representative thermal desorptpwd
ion curves of $HD$ from olivine;
exposure to $H$ and $D$ for 1, 1.5, and 2 minutes (top to bottom) at 5 K; heating rate: $\sim 1 K/sec$.}

\figcaption{ $HD$ desorption rate from olivine vs. exposure. Notice the
different units in the abscissa in the two panels. (Line: fit - see text)}

\figcaption{Desorption yield of $HD$ from olivine after adsorption of $H$
and $D$ at the indicated temperatures. Exposure: 1 minute (Line: guide to the eye).}

\figcaption{$HD$ recombination efficiency $r$ vs. exposure time at $\sim 6
K$. (Line: guide to the eye).}
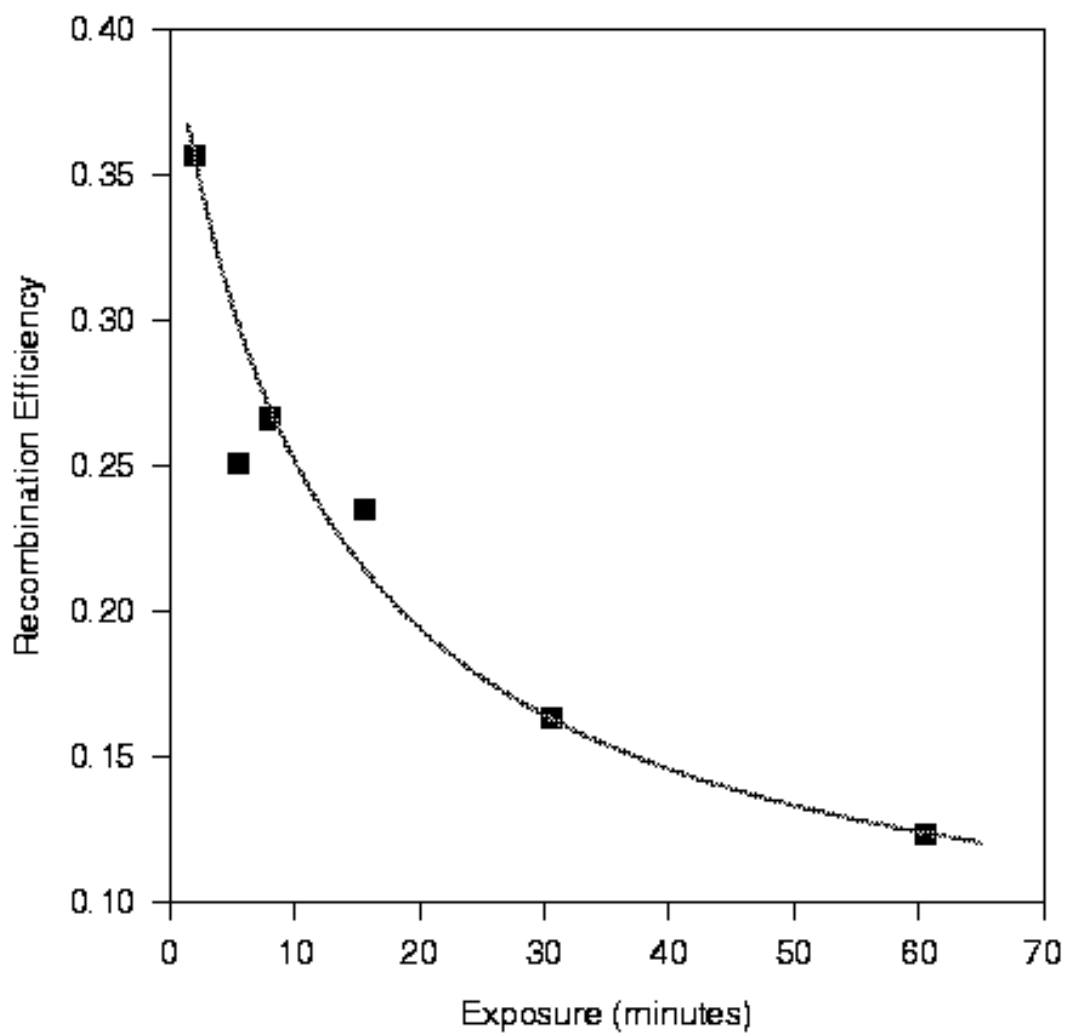
\end{document}